\documentclass[aps,prd,twocolumn,preprintnumbers,showpacs,superscriptaddress,nofootinbib]{revtex4-1}

\IfFileExists{srcltx.sty}{\usepackage[active]{srcltx}}

\usepackage{graphics}  % needed for figures
\usepackage{epsfig}
\usepackage[usenames]{color}
 \usepackage{verbatim}
\IfFileExists{srcltx.sty}{\usepackage[active]{srcltx}}

\begin{document}

\title{TeV gamma rays from blazars beyond z=1?}

\author{Felix Aharonian}
\affiliation{School of Cosmic Physics, Dublin Institute for Advanced Studies, 31 Fitzwilliam Place, Dublin 2, Ireland}
\affiliation{Max Planck Institute f\"ur Kernphysik, 67117 Heidelberg, Germany}
\author{Warren Essey}
\affiliation{Department of Physics and Astronomy, University of California,Los Angeles, CA 90095-1547, USA}
\author{Alexander Kusenko}
\affiliation{Department of Physics and Astronomy, University of California,Los Angeles, CA 90095-1547, USA}
\affiliation{Kavli Institute for the Physics and Mathematics of the Universe, University of Tokyo,
Kashiwa, Chiba 277-8568, Japan}
\author{Anton Prosekin}\thanks{Fellow of the International Max Planck Research
     School for Astronomy and Cosmic Physics at
                    the University of Heidelberg (IMPRS-HD)}
\affiliation{Max Planck Institute f\"ur Kernphysik, Heidelberg, Germany}

\begin{abstract} At TeV energies, the gamma-ray horizon of the universe
is limited to redshifts {$z \ll 1$}, and, therefore,  any observation of TeV radiation from a source  
located beyond  {$z=1$} would call for a revision of the standard paradigm. While robust observational 
evidence for TeV sources at redshifts $z \geq 1$ is lacking at present, the growing number of TeV blazars 
with redshifts as large as  $z \simeq 0.5$ suggests the possibility that the standard blazar models 
may have to be reconsidered.  We  show that  TeV gamma rays can be observed even from a source at $z \geq 1$, 
if the observed gamma rays are secondary photons produced  in interactions of  
high-energy protons originating from the blazar  jet and propagating over 
cosmological distances almost rectilinearly.   This mechanism was initially proposed as a possible 
explanation for the TeV gamma rays observed from blazars with redshifts $z \sim 0.2$, 
for which some other explanations were possible.   For TeV gamma-ray radiation detected from a blazar with  $z\geq 1$,  
this model would   provide  the only viable interpretation consistent with conventional  physics.  
It would also have far-reaching astronomical and cosmological ramifications.  In particular,  
this interpretation would imply that extragalactic magnetic fields along the line of sight are very weak,  
in the range {$ 10^{-17}\, {\rm G} <B < 10^{-14}\, {\rm G}$}, assuming random fields 
with a correlation length of 1~Mpc, and that acceleration of  
$E \geq 10^{17} \ \rm eV$ protons in the jets of active galactic nuclei can be very effective. 
\end{abstract}
\maketitle 

\section{Introduction}
Recent observations of active galactic nuclei with ground-based gamma-ray detectors 
show growing evidence of very high energy (VHE)  gamma-ray emission from blazars with 
redshifts well beyond $z=0.1$.  In this paper we examine the question of whether TeV blazars 
can be observed from even larger redshifts, $z \geq 1$.  Although primary TeV gamma rays produced at the source 
are absorbed by extragalactic background light (EBL), we will show that it is possible to 
observe such distant blazars as point sources due to secondary photons generated along the line of sight 
by cosmic rays accelerated in the source. 

To a large  extent, the observations of blazars with $z> 0.1$  came as a surprise, in view of the severe absorption 
of such energetic gamma rays in the EBL. One of the obvious 
implications of these observations is the unusually hard (for gamma-ray sources)  
intrinsic gamma-ray spectra.  Remarkably,  the {\it observed} energy 
spectra of these objects  in the very high energy band are, in fact, 
very  steep,  with photon indices  $\Gamma \geq 3.5$. However,  after the 
correction for the expected intergalactic absorption 
(i.e. multiplying the {\it observed} spectra to the factor of  $\exp{[\tau(z,E)]}$, where   $\tau(z,E)$ 
is the optical depth of gamma rays of energy $E$ emitted by a source of redshift $z$),  
the intrinsic (source) spectra  appear to be very hard with a photon index 
$\Gamma_{\rm s} \leq 1.5$.  Postulating  that in standard scenarios the gamma-ray production spectra 
cannot be harder than $E^{-1.5}$, it was claimed that the EBL must be quite low, based on the  
observations of  blazars H~2356--309 ($z=0.165$) and 1ES~1101--232 ($z = 0.186$) 
by the HESS collaboration \cite{Aharonian:2005gh}. 
The derived upper limits  appeared to be  rather close to the lower limits on EBL 
set by the integrated light of resolved galaxies. Recent  phenomenological  and theoretical studies 
(e.g., Refs.~\cite{Franceschini,Gilmore2012}) also favor the models of EBL  which are 
close to the limit derived from the galaxy counts (for a recent review see Ref.~\cite{Costamante2012}). 
This implies that further decrease in the level EBL is practically impossible, thus a detection of TeV 
gamma rays from more distant objects would call for  new approaches to explain or avoid the extremely hard intrinsic gamma-ray spectra. 

The  proposed nonstandard  astrophysical scenarios include  models with very hard gamma-ray production spectra  
due to some specific shapes of energy distributions of the parent relativistic 
electrons --  either a power law  with a high low-energy cutoff or a narrow, e.g., Maxwellian-type distribution.  While the  
synchrotron-self-Compton (SSC) models allow  the hardest possible gamma-ray 
spectrum with the photon index $\Gamma=2/3$ \cite{Tavecchio2009,Lefa2011}, the external Compton 
(EC) models can provide gamma-ray spectrum with $\Gamma=1$ \cite{Lefa2011}. Within these models one can explain the gamma-ray emission  of the blazar 1ES~229+200 at $z=0.139$ with the  
spectrum extending up to several TeV \cite{HESS0229} and sub-TeV gamma-ray emission from 3C~279 at $z=0.536$ \cite{Magic:3C279}  ($\Gamma_s \sim 1$).  Formally, much harder spectra can be expected in the case of Comptonization of an ultrarelativistic  outflow \cite{ATPcascade}, in analogy with the cold electron-positron winds in pulsars~\cite{BAh2000}.  Although it is not clear how the   ultrarelativistic  MHD outflows could form in active galactic nuclei (AGN) with a bulk motion Lorentz factor $\gamma \sim 10^6$, such a scenario, leading to the Klein-Nishina  gamma-ray line-type emission~\cite{AKM2012}, cannot be excluded {\em ab initio}. Further hardening of the initial (production) gamma-ray spectra can be realized due to  the internal $\gamma-\gamma$ absorption  inside the source \cite{AKC2008,Zachar2011}. 
Under certain conditions, this process may lead to an arbitrary hardening of the original production spectrum of gamma rays. 

Thus, the failure  of  ``standard" models to reproduce the extremely hard intrinsic  
gamma-ray spectra is likely to be due to the lack of proper treatment of the complexity of nonthermal processes in blazars, 
rather than a need for new physics.    However, the situation  is dramatically  different in the case of blazars with redshift $z \geq 1$. In this case the drastic increase in the optical depth for gamma rays with energy above several hundred GeV implies severe absorption (optical depth $\tau \gg 1$),  which translates into  unrealistic energy budget requirements (even after reduction of the intrinsic gamma-ray luminosity by many orders of magnitude due to the Doppler boosting).  In this case, more dramatic proposals  including   violation of Lorentz invariance~\cite{Kifune99,SteckerGlashow,JacobPiran} 
or  ''exotic`` interactions involving hypothetical axion-like particles~\cite{De_Angelis:2007dy,Simet:2007sa}
are justified.  Despite the very different nature of these approaches, their main objective is the same -- to avoid 
severe  intergalactic absorption of gamma rays due to photon-photon pair production at interactions with EBL.  This feat was accomplished  either by means of big modifications  in the cross-sections,  or by 
assuming gamma-ray oscillations into some weakly interacting particles during their propagation 
through the intergalactic magnetic fields (IGMFs), e.g., via the photon mixing with an axion-like particle.  
Alternatively, the apparent transparency of the intergalactic medium to VHE gamma rays can be increased if 
the observed TeV radiation from blazars is secondary, i.e., if it is formed
in the development of electron-photon cascades in the intergalactic medium initiated by primary 
gamma rays \cite{ATPcascade}. This assumption can, indeed, help us to increase the {\it effective} 
mean free path of VHE gamma rays, and thus weaken the absorption  of gamma rays from 
nearby blazars, such as Mkn~501 \cite{ATPcascade,ATaylor}. However,  
for cosmologically distant objects the effect is almost  negligible because the ''enhanced`` mean free path of gamma rays is 
still much smaller than the distance to the source.   

A modification of this scenario can  explain TeV signals from objects beyond $z=1$ if one assumes that the primary particles initiating the intergalactic cascades  are not gamma rays, but protons with energies  $10^{17}-10^{19}$~eV~\cite{Essey:2009zg,Essey:2009ju,Essey:2010er,Murase:2011cy,Razzaque:2011jc,Essey:2010nd,Essey:2011wv,Prosekin:2012ne}.  
AGN are a likely source of very high energy cosmic rays~\cite{Biermann:1987ep,Aharonian:2001dm}.  High-energy protons can travel cosmological distances and can effectively generate secondary gamma rays along their trajectories.  Secondary gamma rays are produced in interactions of protons with 2.7~K cosmic microwave background radiation (CMBR) and with EBL.

\section{Rectilinear propagation and deflections}
Secondary photons from proton induced cascades point back to the source if the the proton deflections are small~\cite{Aharonian:2001dm}. 
Rectilinear propagation of protons is possible along a line of sight which does 
not cross any galaxies, clusters of galaxies, because their magnetic fields  
would cause a significant deflection.  In addition, IGMFs can cause deflections in the voids, where the fields can be as low as 
$10^{-30}$~G~\cite{magn_fields,astro-ph/0210095}, but the analysis of blazar spectra including cosmic rays and secondary photons 
points to a range from 0.01 to 30 femtogauss~\cite{Essey:2010nd}.  As long as IGMFs are smaller than a femtogauss, they do not affect the point images of blazars.  It remains to show that a typical line of sight does not cross a galaxy, cluster, etc.  The mean rectilinear propagation length for protons reaching us from a distant source was discussed in Ref.~\cite{astro-ph/0210095}.   Given homogeneity of the large-scale structure at 
large redshifts, this distance can be estimated as the mean free path of a 
proton in a volume filled with density $n$ of uniformly distributed scatterers, each of which 
has a size $R$~\cite{astro-ph/0210095}.  A typical distance the proton passes 
without encountering a scatterer is $L \sim 1/(\pi R^2 n)$.  One can estimate this distance for galaxies,
clusters, etc., and adopt a constraint based on the minimal distance $L_{\rm min}$.  Sources at distances much larger than 
$L_{\rm min}$ should not be seen as point sources of secondary photons.  It turns out that the strongest 
limit comes from galaxy clusters~\cite{astro-ph/0210095}: 
\begin{equation}
L_{\rm min} \sim 1/(\pi R^2 n)  \sim (1-5) \times 10^3 {\rm Mpc}. 
\end{equation}
This distance is large enough for a random source at $z\ge 1$ to be seen with no obstruction by  
a cluster, or a galaxy~\cite{astro-ph/0210095}.   
Thus, the protons of relevant energies propagate rectilinearly, assuming the IGMFs are small.  

\begin{figure}[ht!]
  \begin{center}
\includegraphics[width=0.5\textwidth]{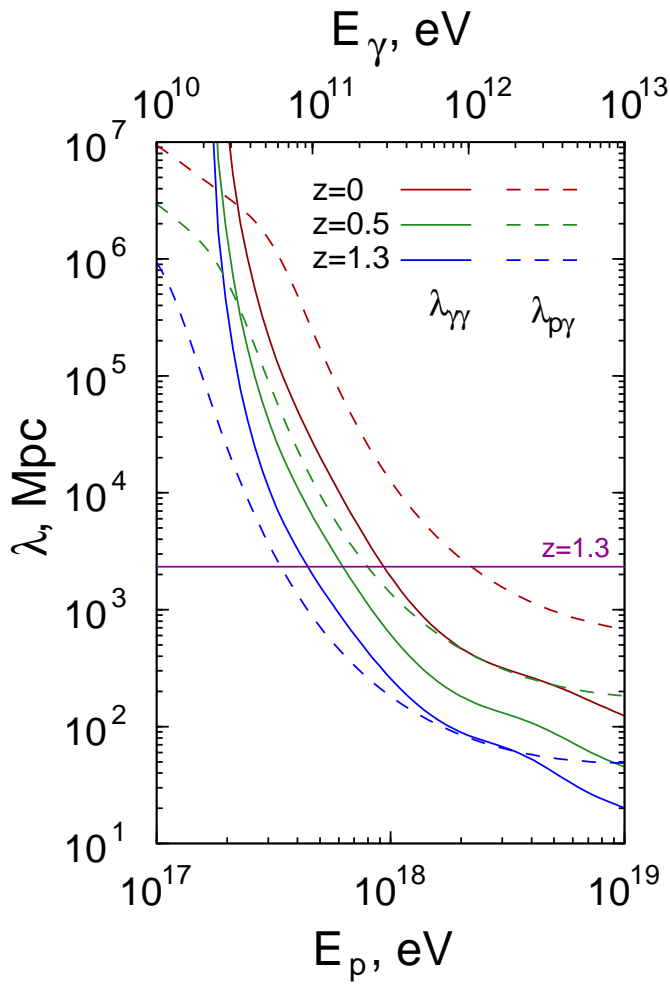}
\caption{The mean free paths of photons and protons as a function of energy and the source redshift. 
The calculations are based on the formalism developed in Ref.~\cite{Prosekin:2011}.  
The gamma-ray absorption mean free path  
$\lambda_{\gamma \gamma}$ is shown for the EBL model of Ref.~\cite{Franceschini}. 
 \label{meanfreepaths}}
    \end{center}
\end{figure}

If IGMFs  on cosmological distance scales 
are smaller than $10^{-15}$G, the protons propagate almost rectilinearly, and they carry some significant 
energy into the last, most important for us segment  of their trajectory determined by the condition 
$l \leq \lambda_{\rm \gamma, eff}$, where $\lambda_{\rm \gamma, eff}$ is the
{\it effective} mean free path of gamma rays. The secondary electron-positron pairs produced with an average energy of 
$(m_{\rm e}/m_{\rm p}) E_{\rm p} \sim 10^{15} \rm eV$ initiate electromagnetic electron-photon 
cascades  supported by the inverse Compton (IC) scattering of electrons  on  CMBR 
and photon-photon pair production of gamma rays interacting with EBL and CMBR. 
As long as the magnetic field is as small as is required to avoid the smearing of point sources, 
the cascade develops with an extremely high efficiency. Therefore, the gamma-ray zone is determined 
by the condition that $\lambda_{\rm \gamma, eff}$ be larger (typically, by a factor of 2 or 3) 
than the gamma-ray absorption mean free path  $\lambda_{\gamma \gamma}$ shown in 
Fig.\ref{meanfreepaths}.  
%
%                             fig1

Our analysis so far (and that of Berezinsky {\em et al}.~\cite{astro-ph/0210095}) left out the filaments between the clusters.  Their size, volume filling factor, and geometry are uncertain, and observations provide only the upper limits.  Models can accommodate a variety of field strengths in these filaments~\cite{Ryu}. If nanogauss fields exist in large, numerous filaments, 
and if the line of sight passes through one or more filaments, the signal strength is reduced, as discussed in Ref.~\cite{Murase:2011cy}.

\section{Energy requirements}
\begin{figure*}[ht]
% \vspace{30mm}
  \begin{center}
      \includegraphics[width=0.45\textwidth]{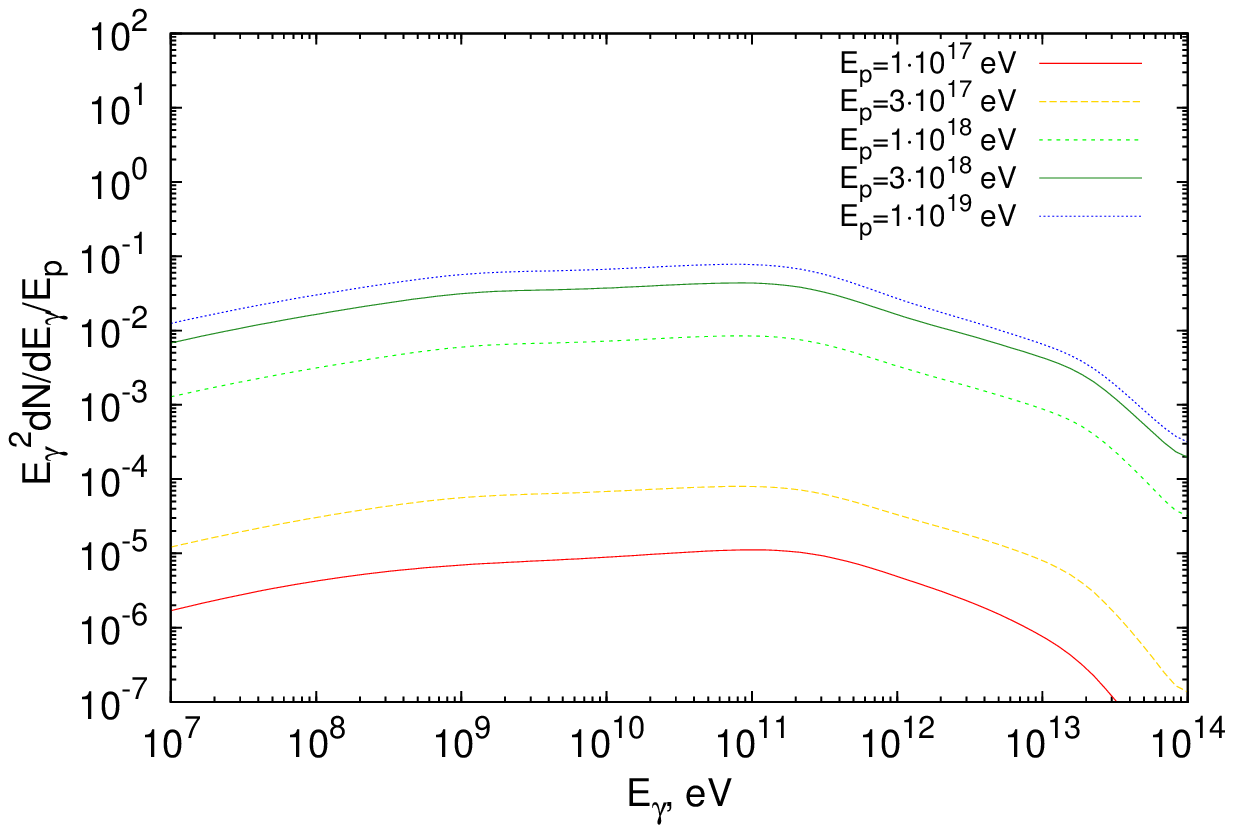}
      \includegraphics[width=0.45\textwidth]{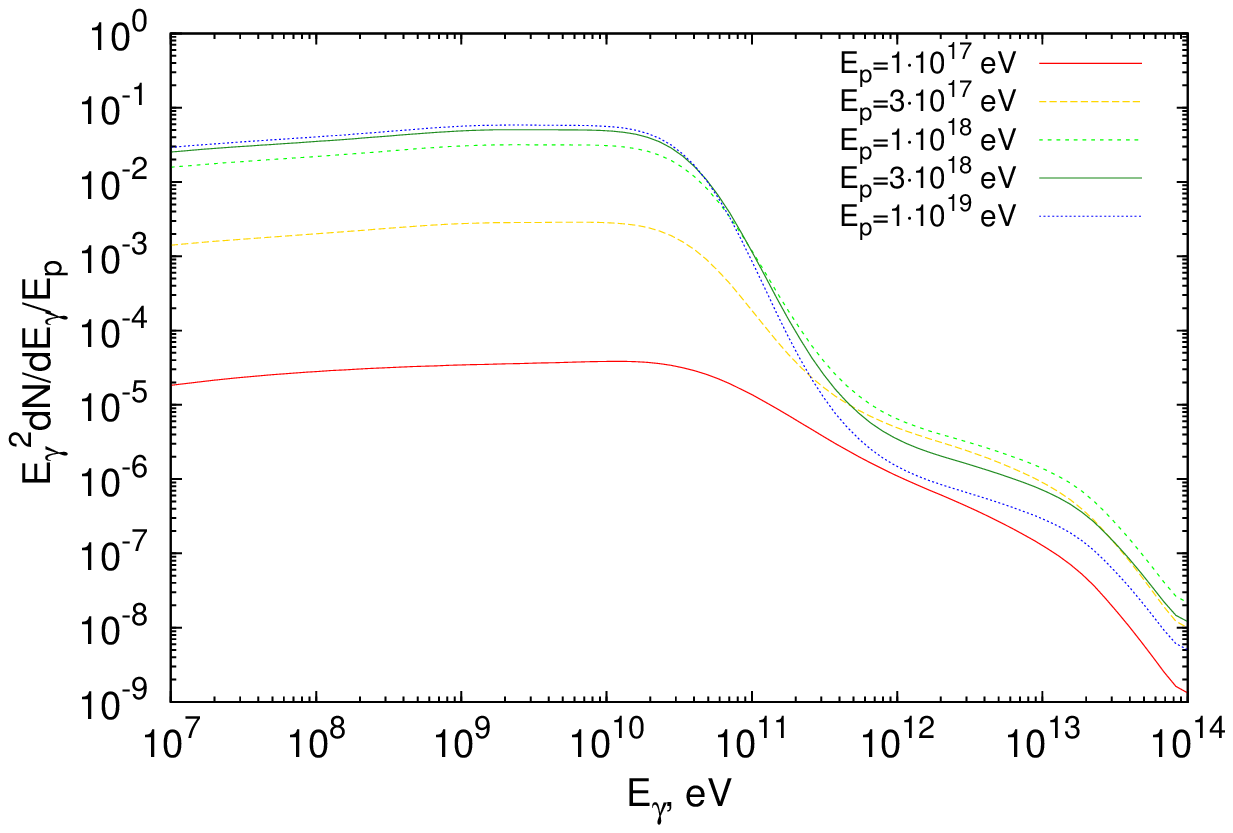}
\caption{The energy spectra  of secondary gamma rays produced by protons of different energies emitted from a source at $z=0.2$ (left panel) and $z=1.3$ (right panel). The curves  are normalized to the 
proton energy, hence, they show the differential 
efficiency of the energy transfer from protons to gamma rays. It is assumed that the intergalactic magnetic field $B=0$. 
 \label{efficiency}}
    \end{center}
\end{figure*}

The efficiency of this  scenario depends on the  energy of primary protons and the size of the {\em gamma-ray transparency zone}.   
It is approximately determined by the fraction of the 
proton energy released in $e^+e^-$ pairs inside the gamma-ray transparency zone, at distances less than  
$\lambda_{\gamma, \rm eff}$ from the observer. Obviously, in the case of a broad energy distribution of 
protons, the main contribution to the gamma-ray flux comes from some energy range in which the proton mean free path 
is comparable to the distance to the source: $d=\lambda_{\rm p \gamma}(E,z=0)$.  
 In the case of nearby objects with $z \ll 1$, the corresponding energy $E^\ast$ can be  found from Fig.~\ref{meanfreepaths} as
the point where the distance to the source 
is equal the mean free path of protons  at the present epoch, 
$d=\lambda_{\rm p \gamma}(E^\ast,z=0)$.
The  contributions of protons with lower or higher  energies would be significantly smaller. For lower energies, 
the interaction probability is too small, while, for higher energies, the energy losses outside the 
gamma-ray transparency zone are too large.  However, in the case of cosmologically 
distant objects, such a simple argument does not work because of very 
strong dependency of the proton's mean free path on both the energy and the redshift.  
It appears that, independent of the initial energy,  only the low-energy protons with $E \sim 10^{17} \ \rm eV$ enter the {\em gamma-ray transparency zone}. This dramatically reduces the efficiency of  production and transport of VHE gamma rays to the observer. At the same time,  the efficiencies  for gamma rays, the mean free paths of which are comparable to the distance to the source, remain high. This is the case for GeV gamma rays from cosmologically 
distant, $z \geq 1$, objects and for VHE gamma rays from small-$z$ objects. This can be seen from Fig.\ref{efficiency}, 
where we show the spectral energy distribution (SED) of gamma rays  normalized to the initial energy of the proton. 
The curves are calculated for two redshifts, $z=0.2$ and $z=1.3$,  and for several different proton energies.    
%
%                               fig2
%
%                                   Fig3 
\begin{figure*}[ht!]
\begin{center}
\mbox{\includegraphics[width=0.45\textwidth]{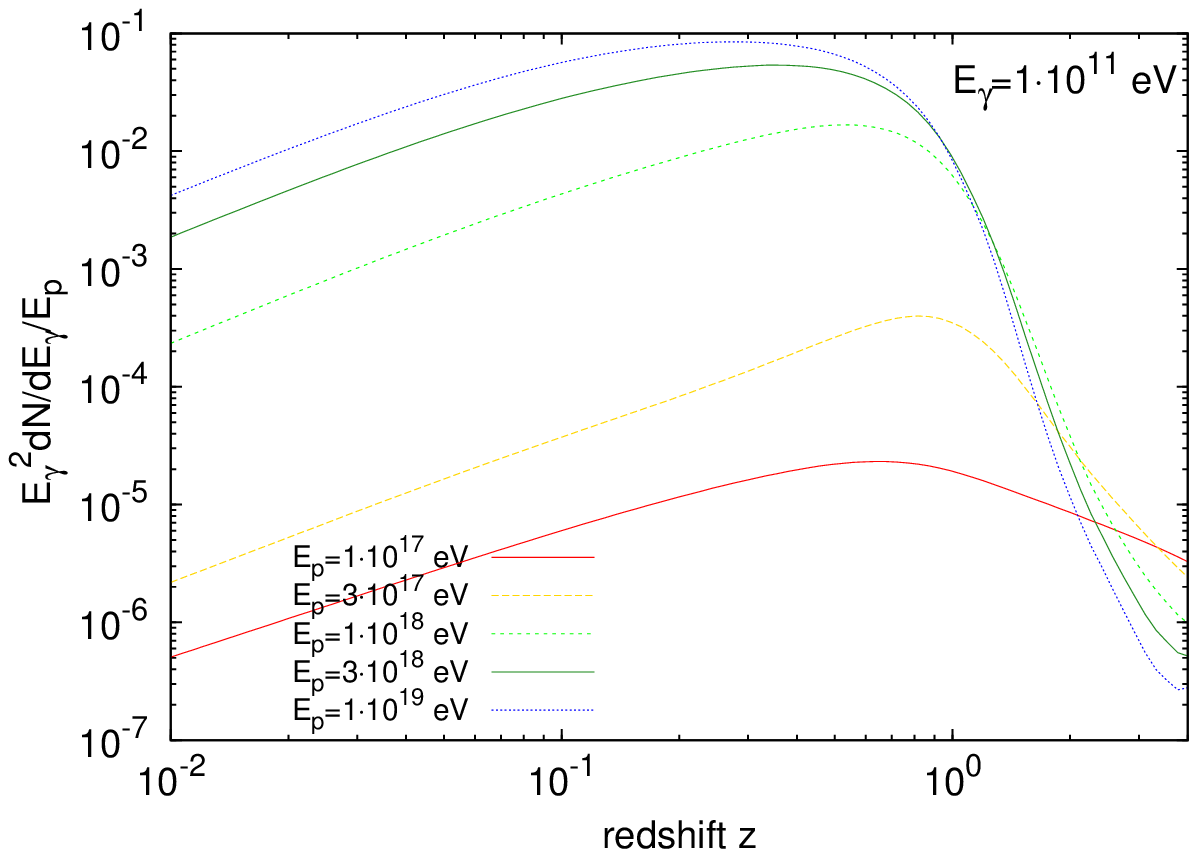}
\includegraphics[width=0.45\textwidth]{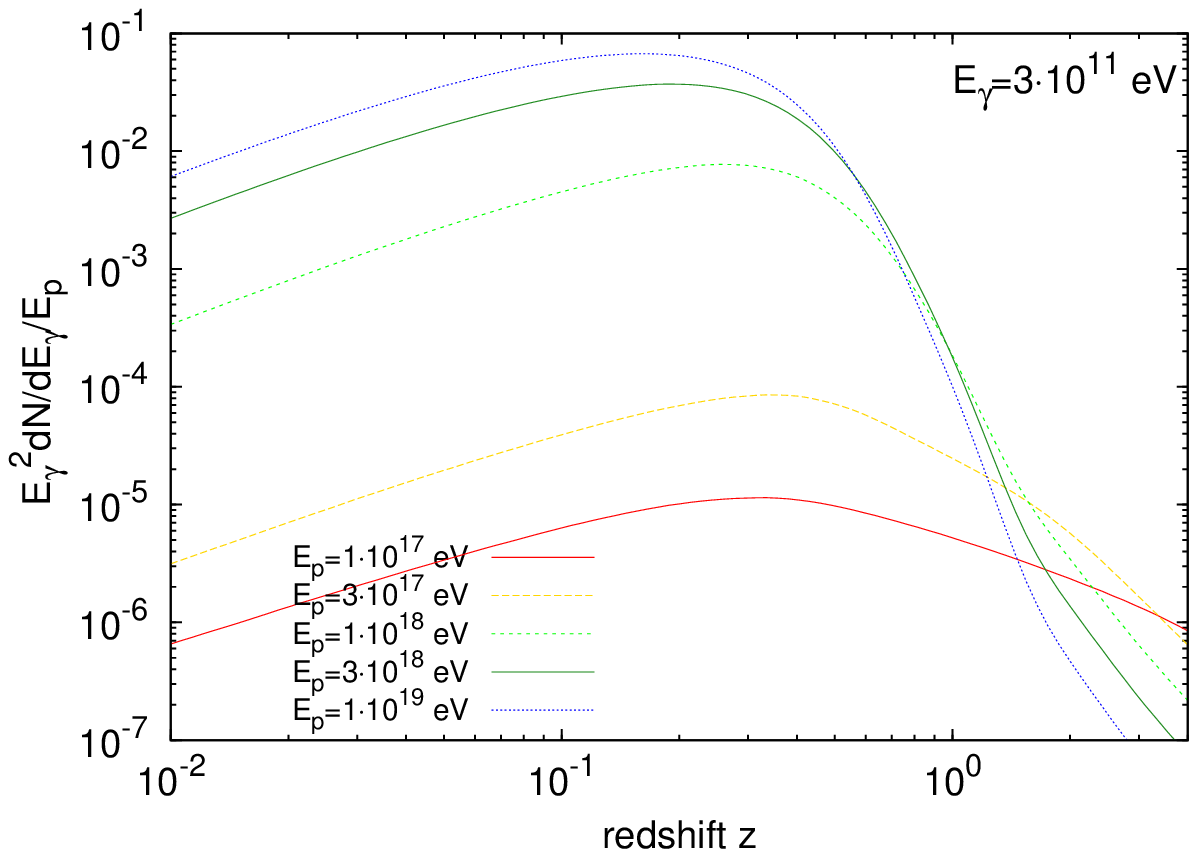}}
\mbox{\includegraphics[width=0.45\textwidth]{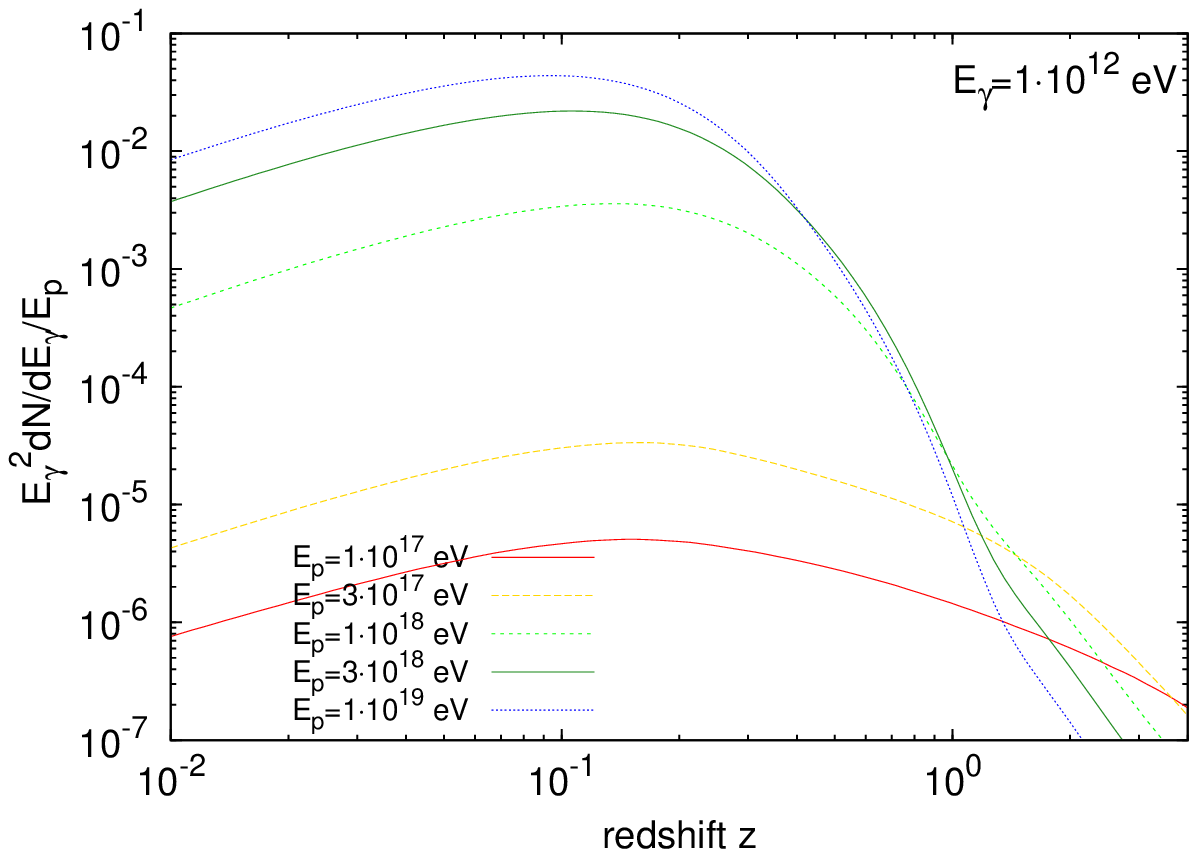}
\includegraphics[width=0.45\textwidth]{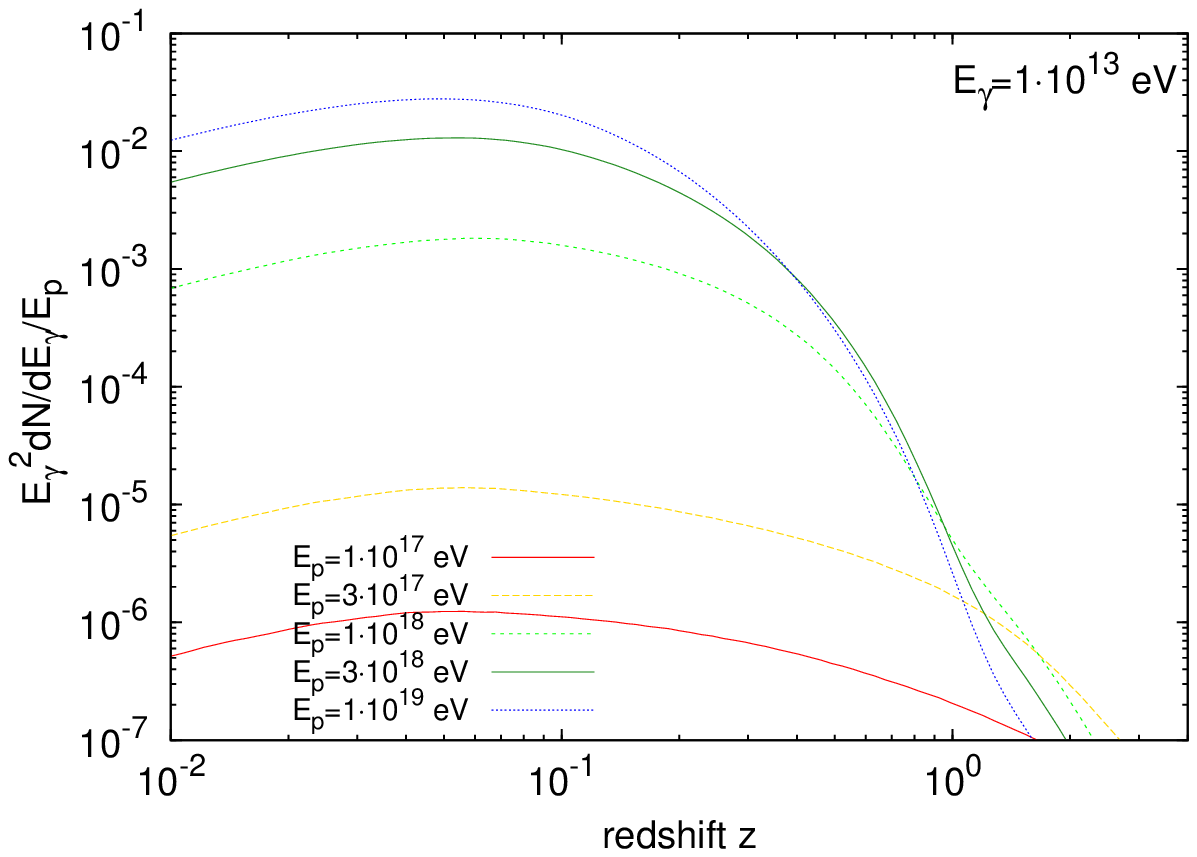}}
\caption{The differential 
efficiency of the energy transfer from protons to gamma rays as a function of 
the redshift of the cosmic-rays source for different initial 
energies  $E_p$ of the monoenergetic proton beam.
\label{diffeff1}}
\end{center}
\end{figure*}

In Fig.~\ref{diffeff1} we show the dependence of the efficiency of energy transfer 
on the redshift of the source. It is determined by the character of evolution of 
radiation fields  with $z$.
While  the energy density of CMBR monotonically decreases with $z$, namely 
$w_{\rm CMBR} \propto (1+z)^4$, the dependence of the  density of EBL on $z$ is more complex and 
uncertain.  
For small redshifts,  the density of EBL increases with $z$, but  
at redshifts corresponding to the epochs before the maximum of the 
galaxy formation rate ($z \sim 2$), the density of EBL is contributed only by the first stars, therefore  
it drops at large redshifts.   Correspondingly, the probability of gamma rays to reach    
the observer has a nonlinear dependence on the energy of protons and the source redshift. 
Depending on the energy of gamma rays,  the efficiency reaches  its maximum at 
intermediate  redshifts, $z \sim 0.1 - 0.3$.  We note that at $z \sim 0.1$,   the efficiency  could be 
rather high (greater than 1\%) even at 10~TeV.  Therefore, the contribution of this channel  to the quiescent  component of 
VHE radiation from nearby blazars  can be quite significant. At large redshifts, $z \geq 1$, the efficiency at TeV energies  drops dramatically, and it does not exceed $10^{-5}$ at $z=1$. Yet, even with such a small efficiency, 
one can expect TeV gamma rays from sources with $z \sim 1$, provided that 
the parent protons leave the blazar in a narrow beam.  In contrast, TeV 
gamma rays emitted directly  by the source at $z \geq 1$ suffer severe absorption, 
thus only a negligible fraction can survive and reach the observer. 
%
%                                   Fig4
\begin{figure*}[ht!]
  \begin{center}
       \includegraphics[width=\textwidth]{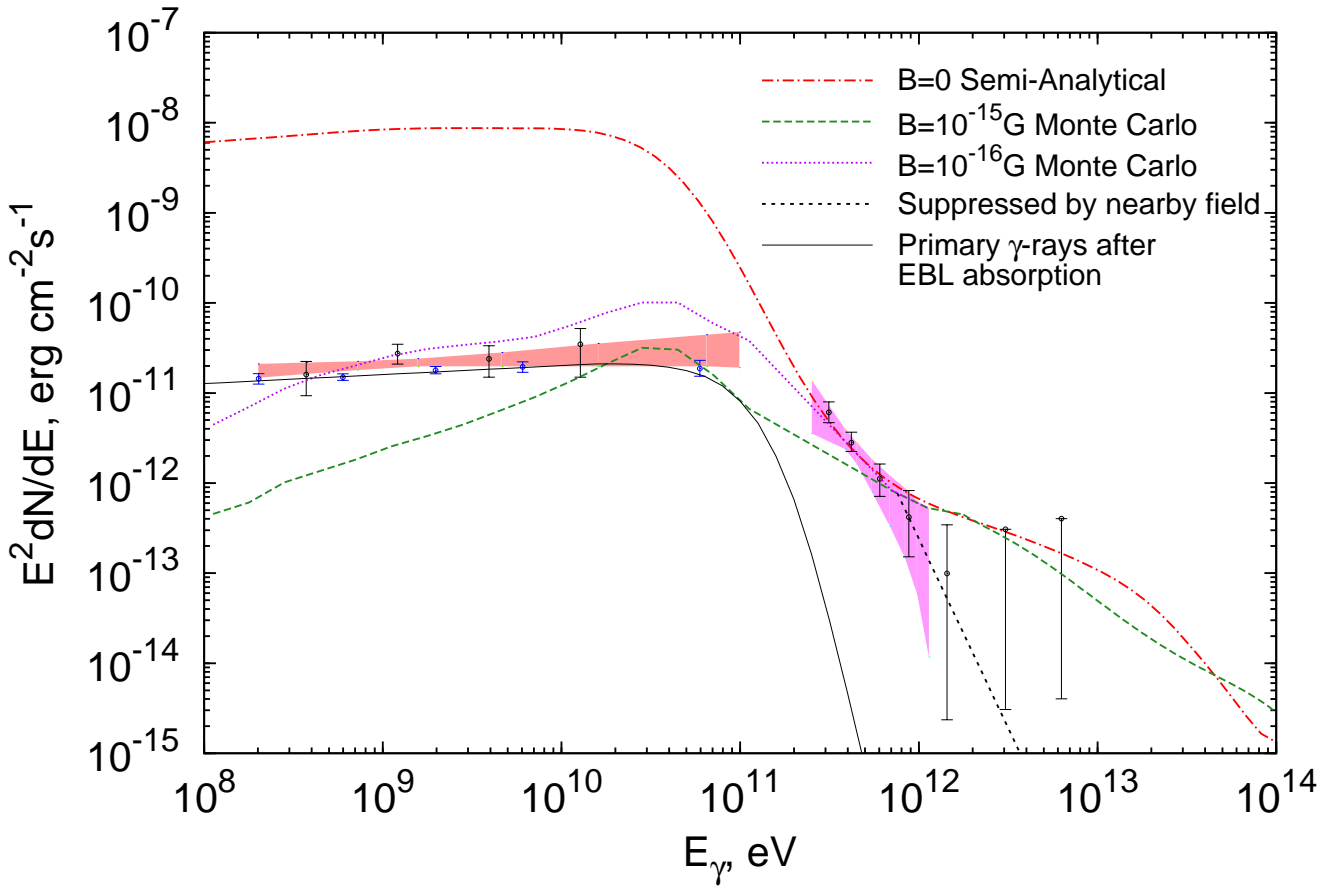}
\caption{Spectra of secondary gamma rays produced by protons from a source at $z=1.3$, 
calculated using semianalytical and Monte Carlo techniques. All theoretical curves are normalized to the observed flux around 1~TeV. 
The { Fermi} LAT data are shown according to 1LAC catalog~\cite{Fermi_LAT} (smaller error bars), and according to Ref.~\cite{Prandini:2011du} (large error bars).  
The data above 0.1~TeV are from HESS~\cite{Zech:2011ym}.
The semianalytical calculations correspond to the magnetic field $B=0$ and protons 
injected with $E_p^{-2}$ type energy spectrum  in the energy interval   $E_p= 10^{17}- 10^{18}$~eV.   Monte Carlo results for 
the secondary spectrum from protons with a high energy cutoff of $10^{19}$~eV  are shown for IGMF $B=10^{-16}$~G and $B=10^{-15}$~G. 
The effect of significantly enhanced magnetic field  within  $D  \stackrel{_<}{_\sim} 100$~Mpc 
of the observer is  shown for illustration of a possible suppression of the spectrum above 1~TeV.  
Also shown is the spectrum from a pure-gamma (no cosmic rays) source with injection spectrum $E_\gamma^{-2}$, after intergalactic absorption for the 
EBL model of Ref.~\cite{Franceschini}.
\label{fig:spectrum}}
    \end{center}
\end{figure*}

Indeed,  for gamma rays with energy in excess of  several hundred GeV arriving from a source at  $z=1$, the optical depth is very large, $\tau_{\gamma \gamma} \sim 10 $,  for any realistic model of EBL.  VHE gamma rays cannot survive the severe intergalactic absorption (see Fig.~\ref{fig:spectrum}).  This could be relevant for TeV gamma-ray emission from the blazar  PKS~0447-439~\cite{Zech:2011ym}, given the large redshift of the source  $z \geq 1.126$, as claimed in Ref.~\cite{Landt:2012it}. However, recently two independent groups~\cite{Pita:2012gv} challenged the interpretation of the redshift measurements of Ref.~\cite{Landt:2012it}. Thus, the redshift  of PKS~0447-439 remains uncertain.

\section{Case study: a blazar at $z=1.3$}
Regardless of the observational status of PKS~0447-439 redshift, it is important to understand whether secondary gamma rays can be detected from a source at a large redshift.  Therefore, we use PKS~0447-439 as a case study for this more general question, assuming it has a redshift 
$z \approx 1.3$, as claimed in Ref.~\cite{Landt:2012it}.  The analysis presented below should be viewed as a  
methodological study whose goal is to demonstrate that the model does allow TeV blazars at redshifts $z \geq 1$ 
to be observed, and that neither a dramatic revision of high-energy processes in blazars, nor new nonstandard interactions of gamma rays are necessary.

Cosmic-ray protons with energies  $E \leq 10^{18}$~eV do  not lose a significant part of their energy to interactions with the  background photons, and,  as long as  the IGMFs   are  very weak, the  protons can provide an effective transport 
of the energy over a large (cosmological) distance toward the observer. 
Cosmic ray interactions with CMBR and EBL, via the Bethe-Heitler pair production  
$p\gamma  \rightarrow p e^+ e^-$ and the photomeson reactions 
$p+\gamma_b  \rightarrow  p + \pi^0$,  initiate electromagnetic cascades. The resulting secondary VHE gamma rays are observed as  arriving from a point source, provided that the broadening of both the proton beam and the cascade electrons due to the 
deflections  in IGMFs  does not exceed the point spread function of the detector.
In the case of detection of VHE gamma rays from PKS~0447-439  by the HESS telescope array~\cite{Zech:2011ym},   $\theta_{\rm p}, \theta_{\rm cas} \leq  3$~arcmin.   While the broadening of the proton beam takes place over the entire path of protons from the source to the observer (zone 1),  the diffusion  of electrons in the {\em transparency zone} (zone 2) is the most important factor for the broadening of the cascade emission. Therefore,  strictly speaking,  one should distinguish between the magnetic fields in these two zones, $B_1$ and $B_2$, respectively. The corresponding deflection angles  are~\cite{theta1}
\begin{equation}
 \theta_{\rm p} \approx 0.05 \ {\rm arcmin} \, \left (\frac{10^{18} \rm eV}{E_p} \right ) 
\left (\frac{B_1}{10^{-15} {\rm G}} \right )  \left (\frac{L}{\rm Mpc} 
\frac{d}{\rm Gpc} \right)^{1/2} 
\end{equation}
and 
\begin{equation}
\theta_{\rm cas} \approx 3.8 \ {\rm arcmin} \, \left ( \frac{10^{12} \rm  eV}{E_\gamma} \right ) \left (\frac{B_2}{10^{-15} \rm G} \right ),
\end{equation}
%(see e.g. Ref. \cite{theta2}), 
where  $L$ is the coherence length, and $d$ is luminosity distance.  One can see that, for comparable 
strengths of  magnetic fields in two zones, the angular broadening 
is mainly due to the electron deflections  in the {\em transparency zone}. 
Remarkably, such a deflection depends only on the magnetic field 
$B_2$ and the gamma-ray energy $E_\gamma$.  
Thus,  a detection of an energy-dependent angular broadening of gamma-ray emission from blazars can provide  
a direct measurements of IGMF in a given direction~\cite{Ando:2010rb}.  

The deflections of protons and cascade electrons result in delays of the 
arrival times of the signal. In the two zones defined above, 
\begin{eqnarray}
\Delta \tau_{\rm p} \approx  1.5 \cdot 10^6 \, {\rm s} \ & & \left ( \frac{E_{\rm p}}{10^{18} {\rm eV}} \right )^{-2} 
\left (\frac{B}{10^{-15} \rm G} \right )^2 \times \nonumber 
\\ & & \times \left ( \frac{L}{1\, \rm Mpc} \right) \left( \frac{d}{1 \rm Gpc} \right)^2 
\end{eqnarray}
and 
\begin{equation}
 \Delta \tau_\gamma \approx 1.3 \cdot 10^6 {\rm s}  \left (  \frac{E_\gamma}{10^{12}  \  \rm eV} \right)^{-5/2} 
\left ( \frac{B}{10^{-15} \rm G}\right )^{2}.
\end{equation}
One can see that, for 
$B_{1} \sim B_{2} \sim 10^{-15} \rm \ G$, any time structure 
in the initial signal of  $10^{18}$eV protons on time scales of the order of a month or shorter 
are smeared out.  Conversely, the interpretation of  a variable 
VHE gamma-ray signal on time scales less than 1~month, in the framework of this model,  would require magnetic field in both zones to be significantly weaker than $10^{-15} \  \rm G$. On the other hand, even for such small magnetic fields, the gamma-ray signals at GeV energies should be stable  on time scales of tens of years. 
 
Finally, a distinct feature of the proposed model is the spectral shape of gamma radiation. 
For relatively nearby sources, $z \ll 1$, the gamma-ray spectrum is flat, with a modest  maximum around $10^{11}$~eV.  
For  cosmologically distant sources with $z \geq 1$,  the spectrum is steep in the sub-TeV part of the spectrum  (down to 10 GeV), 
with a tendency of noticeable  hardening above 1 TeV (see Fig.~\ref{efficiency}). Remarkably, 
the spectrum effectively extends to 10~TeV and higher energies even for cosmologically distant objects.   However, a cutoff in the spectrum below a TeV energy cannot be excluded if the magnetic field in the $\approx 100$~Mpc vicinity of the observer  significantly exceeds  $10^{-15} \ \rm G$. 

For a nearby source, the spectral shape of secondary photons is remarkably independent of the details of the proton energy  spectrum~\cite{Essey:2009ju,Essey:2010er}, although the 
efficiency decreases dramatically for the proton energy below $10^{18}$~eV.   For cosmologically distant sources, the shape of the gamma-ray spectrum does depend on the proton energy, especially at $E \leq 10^{18}$~eV.  For a source at $z \geq 1$, the proton energy is transferred to gamma rays with a maximal efficiency if $E \approx 10^{18}$eV.   Therefore, for an arbitrary spectrum of cosmic rays,  the main contribution to secondary gamma rays comes from a relatively narrow energy interval of protons around $10^{18}$~eV.  On the other hand, the gamma-ray spectrum produced  by these protons  in extremely low IGMF ($B \leq 10^{-17}$~G) disagrees  with  the broadband SED of gamma rays  detected by {Fermi} LAT and 
HESS  as shown in Fig.~\ref{fig:spectrum}. This suggests the presence of magnetic fields stronger than $10^{-17}$~G.
In a  stronger  magnetic field,  deflections of the cascade electrons make the gamma-ray  beam at low energies broader.  
The deflected flux does not contribute to a point source, but rather to 
the diffuse extragalactic background radiation.  Meanwhile, VHE gamma rays may be confined  in the initial  narrow beam. This effect is demonstrated in 
 Fig.~\ref{fig:spectrum} which is produced  using the method described in Ref.~\cite{Essey:2010er}.  
 For the IGMF  $B  \geq 10^{-17}$~G, the GeV gamma-ray flux within an angle corresponding to the PSF of HESS, drops by two orders of magnitude to the level detected by { Fermi}  LAT.  The impact on the spectrum of VHE gamma rays is less
pronounced, unless the magnetic field  exceeds $10^{-14}$~G.    

The results presented in  Fig.~\ref{fig:spectrum} show that 
secondary gamma rays can  describe correctly the spectrum of  PKS~0447-439, as long as IGMFs are in the range $10^{-17}{\rm G} < B <10^{-14}{\rm G}$, assuming random fields with a correlation length of 1~Mpc.  This range of IGMF can be narrowed significantly 
in the future angular and temporal studies, leading to a more precise measurement of the magnetic field strengths along the line of sight. For example, detection of variability of VHE emission on timescales less than a few days would imply the values of magnetic fields close to $10^{-17}$~G.  
It is also important to  search for an  unavoidable (in the framework of this model)  broadening of the angular extent of gamma-ray signals from cosmologically distant blazars. The choice of the gamma-ray energy for such studies depends on the magnetic field.  
The  detection of such an effect would be another strong argument in favor of the proposed scenario, and it would  
allow an accurate measurements of IGMFs in different directions. 

\section{Discussion}
One can see from  Fig.~\ref{fig:spectrum} that the energy spectrum of gamma rays is quite stable from several hundred GeV to 10 TeV and beyond.  Although the current  statistics of the results  reported by HESS does not allow  robust  conclusions  regarding the energy spectrum above 1~TeV, the detection  of multi-TeV gamma rays  from PKS~0447-439 as well as from other cosmologically distant blazars  would not be a surprise, but rather a natural consequence of the proposed scenario. However, we note that, if the magnetic field is enhanced in the {\em transparency zone}, i.e.  in the  vicinity of  the observer, 
it could cause a strong suppression  of the gamma-ray flux above some 
energy which can be found  from  the condition 
$\lambda_{\gamma \gamma}(E)=D$.  The impact of this effect on the gamma-ray spectrum detected by an observer 
strongly depends on the linear scale of the enhanced magnetic field, $D$, but not much on the magnetic field itself 
(as long  as the latter is significantly  larger than $10^{-15}$G). For example, for $D \sim 300$~Mpc, the  steepening of the gamma-ray spectrum starts effectively around 1~TeV.  This effect is   illustrated qualitatively in Fig.~\ref{fig:spectrum}.

The isotropic luminosity of the source in  protons 
required to explain the data~\cite{Zech:2011ym},  is in the range  $(1-3)\times10^{50}$~erg/s, depending on the spectrum of protons.  
This is an enormous, but not an unreasonable power, given that the 
actual (intrinsic)  luminosity can be smaller  by several orders of magnitude if the protons are emitted in a small angle. In particular, for  $\Theta = 3^\circ$, the intrinsic luminosity is comparable to the Eddington luminosity of a black hole with a mass $M \sim  10^{9} M_\odot$.   Assuming that only a fraction of the blazar jet energy is transferred to high-energy particles, the jet must operate at a super-Eddington luminosity. 
While it may seem extreme, this suggestion does not contradict the basic principles of accretion, provided that most of the accretion  energy is converted to the kinetic energy of an outflow/jet, rather than to thermal radiation of  the accretion flow. Moreover, there is growing evidence of super-Eddington luminosities  characterizing relativistic outflows in GRBs and in very powerful blazars~\cite{Ghisellini}.   

Finally, we note that the protons emitted by cosmologically distant objects  
are potential contributors  to the  diffuse gamma-ray background.  The total energy deposited into the 
cascades through secondary Bethe-Heitler pair production  does not depend on the orientation of the jet or the beaming angle, 
but only on the injection power of  $\geq 10^{18}$eV protons and on the number of such objects in the universe. 
Generally,  the total  energy flux of gamma rays is fairly independent of the  
strength of the  intergalactic magnetic fields, except for the highest energy part of the gamma-ray spectrum. 
If the contribution of these sources to the diffuse gamma-ray background is dominated by cosmologically 
distant objects, then the development of the proton-induced electron-photon cascades is saturated at large redshifts. 
One should, therefore,  expect a rather steep (strongly attenuated)  spectrum of diffuse gamma rays above 100~GeV.   
However, in the case of very small intergalactic magnetic fields,  the $10^{18}$eV protons can bring significant amount of nonthermal energy to the nearby universe, and thus enhance the diffuse background by TeV photons. Perhaps, this can explain the unexpected  excess of VHE photons in the spectrum of the diffuse gamma-ray background as revealed recently by the { Fermi} LAT data~\cite{Murase:2012}. 
   
\acknowledgments
The work of A.K. was supported in part by the DOE Grant DE-FG03-91ER40662 and by World Premier International Research Center Initiative, MEXT, Japan.

\end{document}